\documentclass[11pt]{article}
\usepackage{amsmath, amssymb, amsfonts}
\usepackage{physics}
\usepackage{hyperref}
\usepackage{bm}
\usepackage{geometry}
\usepackage{float}
\usepackage{authblk}
\usepackage[utf8]{inputenc}
\usepackage[T1]{fontenc}
\usepackage{graphicx}
\usepackage{subcaption}
\title{Adaptation of Quantum Models to Economic Growth Theories}
\author{
  Hugo Spring-Ragain \\
  Centre d'Études Diplomatiques et Stratégiques (CEDS) \\
  \texttt{hugo.springragain@edu.ceds.fr} \\
  \texttt{raguhugo37@gmail.com}
}
\date{\today}

\begin{document}
\maketitle

\begin{abstract}
Traditional economic growth theories, grounded in deterministic and often linear frameworks, fail to adequately capture the inherent uncertainty, non-commutativity, and complex interdependencies of modern economies. This paper proposes a novel approach by transposing fundamental concepts of quantum mechanics---such as superposition, operator algebra, and path integrals---into the realm of macroeconomic modeling. Within this quantum framework, core economic variables (capital, labor, and technological progress) are redefined as non-commuting operators acting on Hilbert spaces, and the state of the economy is represented as a dynamic wave function governed by a time-dependent Hamiltonian. The evolution of this economic wave function follows a generalized Schrödinger equation, developed here through Dyson series and Magnus expansions. We also define a quantum production function as the expected value of a composite operator, capturing the probabilistic nature of economic output. By integrating uncertainty relations analogous to Heisenberg's principle, and modeling economic fluctuations via Langevin dynamics, we extend the model to include dissipation, feedback loops, and non-linear interactions between variables. Finally, a Feynman path integral formalism is constructed to provide an alternative trajectory-based interpretation of economic dynamics. This quantum-inspired framework offers a rigorous and flexible methodology to rethink macroeconomic modeling under radical uncertainty, with potential applications in dynamic policy simulations and innovation-driven growth.
\end{abstract}
\section{Introduction}
Traditional economic growth theory traditionally relies on macroeconomic models in which an economy’s aggregate production is expressed as a deterministic function of the production factors. The most common and simplest model, based on a neoclassical approach, uses an aggregate production function of the Cobb-Douglas type. This functional form relates the production level, denoted $Y(t)$, to two key factors—physical capital $K(t)$ and labor $L(t)$—all modulated by a technological progress factor $A(t)$. The classical scalar model is written as:
\begin{equation}
Y(t) = A(t) K(t)^\alpha L(t)^\beta
\end{equation}
where $\alpha$ and $\beta$ are parameters between 0 and 1, representing the relative share of each factor in the production process. In this representation, $A(t)$ captures the state of technological advancement, $K(t)$ the stock of physical capital (machinery, infrastructure, etc.), and $L(t)$ the available labor. This model, in its exogenous (Solow model) or endogenous (Romer, Lucas, etc.) variants, has helped to explain the fundamental mechanisms of capital accumulation, the role of innovation, and the long-term dynamics of per capita income.

However, it must be acknowledged that despite its success, the scalar model presents limitations when describing the complex and uncertain nature of contemporary economic dynamics. In a classical framework—even when incorporating stochastic shocks—the variables remain commutative and precisely determined at each instant. The order and manner in which factors adjust generally have no fundamental impact on the final macroeconomic outcome. Moreover, economic uncertainty is most often modeled as a simple additive noise without questioning the very structure of the production function.

Yet, in many contexts, economic reality reveals nonlinear interactions, complex interdependencies, and order effects in decision-making that are not easily captured by a strictly classical approach. Investment decisions in capital, the introduction of new technologies, or labor market adjustments are not necessarily interchangeable: introducing technology before accumulating capital may impact the growth trajectory differently than doing it the other way around. These subtleties, as well as the presence of uncertainties more fundamental than mere stochastic variations, invite the exploration of new analytical frameworks.

\section{Transition to a Quantum Model}

In the quantum model presented here, classical economic variables such as capital, labor, and technological progress are transformed into operators acting on Hilbert spaces. Unlike traditional approaches in which these variables are precisely determined, here they are represented by superpositions of states—reflecting the inherent uncertainty of economic dynamics.
\subsection*{Quantum State of the Economy}

The quantum state of the economy at a given instant $t$ is described by a state vector $\ket{\psi(t)}$, which is a superposition of the possible states of $\hat{K}(t)$, $\hat{L}(t)$, and $\hat{A}(t)$:

\begin{equation}
\ket{\psi(t)} = \sum_i \alpha_i(t) \ket{K_i} + \sum_j \beta_j(t) \ket{L_j} + \sum_k \gamma_k(t) \ket{A_k}
\end{equation}

\subsection*{Quantum Operators}

Capital $\hat{K}(t)$ acts on a Hilbert space of capital $\mathcal{H}_K$. Here, capital is no longer a determined quantity but a superposition of different levels of investment:

\begin{equation}
\hat{K}(t) = \sum_i c_i(t) \ket{K_i}
\end{equation}
Labor $\hat{L}(t)$ is an operator in a space $\mathcal{H}_L$ with states $\ket{L_j}$ representing the available labor in different economic sectors:

\begin{equation}
\hat{L}(t) = \sum_j d_j(t) \ket{L_j}
\end{equation}
Technological progress $\hat{A}(t)$ is an operator in a space $\mathcal{H}_A$ representing innovation at various levels:

\begin{equation}
\hat{A}(t) = \sum_k e_k(t) \ket{A_k}
\end{equation}
\subsection*{Quantum State of the Economy}

The quantum state of the economy at a given instant $t$ is described by a state vector $\ket{\psi(t)}$, which is a superposition of the possible states of $\hat{K}(t)$, $\hat{L}(t)$, and $\hat{A}(t)$:

\begin{equation}
\ket{\psi(t)} = \sum_i \alpha_i(t) \ket{K_i} + \sum_j \beta_j(t) \ket{L_j} + \sum_k \gamma_k(t) \ket{A_k}
\end{equation}
\subsection*{Hamiltonian and Dynamic Evolution}

The quantum formalism adopted to model the economy considers the state of the system at time $t$ as a vector $\ket{\psi(t)}$ in a Hilbert space $\mathcal{H}$. The evolution of the system is governed by the Schrödinger equation, which here describes the evolution of the state of the economy:

\begin{equation}
i\hbar \frac{\partial}{\partial t} \ket{\psi(t)} = \hat{H}(t)\ket{\psi(t)}
\end{equation}
The Hamiltonian $\hat{H}(t)$ describes the total energy of the economic system; it is a combination of the operators $\hat{K}(t)$, $\hat{L}(t)$, $\hat{A}(t)$:

\begin{equation}
\hat{H}(t) = \alpha \hat{K}(t) + \beta \hat{L}(t) + \gamma \hat{A}(t)
\end{equation}
Where $\alpha$, $\beta$ and $\gamma$ are real coefficients whose economic interpretation may vary. By analogy with quantum physics, $\hat{H}(t)$ can be seen as an operator representing the “energy value” of the economy, encompassing its capital, labor and technological potential.

Assume then that $\hat{K}(t)$, $\hat{L}(t)$ and $\hat{A}(t)$ are diagonalizable and that we have bases of eigenstates. We introduce: 

\begin{itemize}
\item A basis $\{\ket{K_i}\}$ of $\mathcal{H}_K$, such that $\hat{K}(t)\ket{K_i} = K_i(t)\ket{K_i}$
\item A basis $\{\ket{L_j}\}$ of $\mathcal{H}_L$, such that $\hat{L}(t)\ket{L_j} = L_j(t)\ket{L_j}$
\item A basis $\{\ket{A_k}\}$ of $\mathcal{H}_A$, such that $\hat{A}(t)\ket{A_k} = A_k(t)\ket{A_k}$
\end{itemize}
On the tensor space $\mathcal{H} = \mathcal{H}_K \otimes \mathcal{H}_L \otimes \mathcal{H}_A$, we decompose the Hamiltonian so that:

\begin{equation}
\hat{H}(t) = \alpha \hat{K}(t) \otimes I_L \otimes I_A + \beta I_K \otimes \hat{L}(t) \otimes I_A + \gamma I_K \otimes I_L \otimes \hat{A}(t)
\end{equation}
Where $I_K$, $I_L$, $I_A$ are the identity operators on each of the subspaces. In the tensor basis $\{\ket{K_i L_j A_k}\}$, the matrix $\hat{H}(t)$ is diagonal if $\hat{K}(t)$, $\hat{L}(t)$ and $\hat{A}(t)$ are diagonal in the chosen bases. Then:

\begin{equation}
\hat{H}(t)\ket{K_i L_j A_k} = \left[\alpha K_i + \beta L_j + \gamma A_k(t)\right] \ket{K_i L_j A_k}
\end{equation}
The exact solution of Schrödinger's equation depends on the form of $\hat{H}(t)$. So when $\hat{H}(t)$ depends on time we have: 

\begin{equation}
i\hbar \frac{\partial}{\partial t} \ket{\psi(t)} = \hat{H}(t)\ket{\psi(t)}
\end{equation}
The formal solution is then written:

\begin{equation}
\ket{\psi(t)} = \tau \exp\left(-\frac{i}{\hbar} \int_0^t \hat{H}(t') dt' \right) \ket{\psi(0)}
\end{equation}
Where $\tau$ is our temporal ordering operator. Unlike the case of a time-independent Hamiltonian, we cannot simply factor the exponential and we cannot, in general, reduce the problem to the simple exponentiation of $\hat{H}(t)$. 

To develop the Schrödinger equation of our model we use a Dyson series, a perturbative expansion of the evolution operator, posing:

\begin{equation}
\hat{U}(t,0) := \tau \exp\left(-\frac{i}{\hbar} \int_0^t \hat{H}(t') dt' \right)
\end{equation}
Which we subsequently develop into an infinite series:

\begin{align}
\hat{U}(t,0) &= I - \frac{i}{\hbar} \int_0^t \hat{H}(t_1) dt_1 \notag \\\\
&+ \left(\frac{i}{\hbar}\right)^2 \int_0^t dt_1 \int_0^{t_1} dt_2 \hat{H}(t_1) \hat{H}(t_2) + \cdots
\end{align}
This expansion iterates over time, taking into account that $\hat{H}(t)$ may not commute with itself at different times. With $[\hat{H}(t_1), \hat{H}(t_2)] \neq 0$ for $t_1 \neq t_2$.

This non-commutation prevents the exponential from being directly factorized, so Magnus' formula must be used to rewrite this ordered exponential as a single exponential, an operator $\Omega(t)$ expressing itself as an---a priori---infinite series of nested commutators. Magnus' formula asserts the following equality:

\begin{equation}
\tau \exp\left(-\frac{i}{\hbar} \int_0^t \hat{H}(t') dt' \right) = \exp(\Omega(t))
\end{equation}
With:

\begin{equation}
\Omega(t) = \sum_{n=1}^{\infty} \Omega_n(t)
\end{equation}
The Magnus formula proposes a different way of writing than the Dyson series seen above: instead of having a sum of products $\hat{H}(t_1)\hat{H}(t_2)\ldots$ in the form of a series, we'll look for the unique operator $\Omega(t)$ such that $\hat{U}(t,0) = \exp(\Omega(t))$. However, note that $\Omega(t)$ itself becomes a series in nested commutators of $\hat{H}$ at different times.
The term $\Omega(t)$ then declines into two terms---as well as an optional third---:

\begin{itemize}
  \item $\Omega_1(t)$:
  \begin{equation}
  \Omega_1(t) = -\frac{i}{\hbar} \int_0^t \hat{H}(t_1) dt_1
  \end{equation}
  Thus, this first term is the simple integral of $\hat{H}(t_1)$ between 0 and $t$, up to a factor of $-i/\hbar$. It is close to the principal term obtained if $\hat{H}$ commuted with itself at all times.
  
  \item $\Omega_2(t)$:
  \begin{equation}
  \Omega_2(t) = -\frac{1}{2\hbar^2} \int_0^t dt_1 \int_0^{t_1} dt_2 \left[ \hat{H}(t_1), \hat{H}(t_2) \right]
  \end{equation}
Here we have the first commutator $[\hat{H}(t_1), \hat{H}(t_2)]$. The factor $-1/(2\hbar^2)$ and the order of the integrations (i.e. $0 \leq t_2 \leq t_1 \leq t$) follow from the fact that we correct for the lack of ordering in the exponential $\exp(\Omega)$.

If, by assumption, $[\hat{H}(t_1), \hat{H}(t_2)] = 0$ for all $t_1, t_2$, then $\Omega_2(t)$ will be zero and our development will stop at $\Omega_1(t)$.

  \item $\Omega_3(t)$:
  \begin{multline}
  \Omega_3(t) = -\frac{i}{6\hbar^3} \int_0^t dt_1 \int_0^{t_1} dt_2 \int_0^{t_2} dt_3 \big( [\hat{H}(t_1),[\hat{H}(t_2),\hat{H}(t_3)]] \\ + [\hat{H}(t_3),[\hat{H}(t_2),\hat{H}(t_1)]] \big)
  \end{multline}
The most common form, known as the “standard version,” in fact includes permutation terms $(t_1,t_2,t_3)$.
\end{itemize}
The principle being that each additional order (3,4,5,...) causes increasingly longer nested commutators to appear. Each of the commutators must be integrated over the time domains $[0 \leq t_n \leq \ldots \leq t_2 \leq t_1 \leq t]$.

In practice, $\Omega_n(t)$ contains integrals that can be multiple up to $n$ times and iterated commutators up to $n$ operators $\hat{H}(t_k)$. Our difficulty then lies in carrying over the analysis of $\Omega(t)$, whose series can be difficult to truncate. Symbolically, we can write:

\begin{equation}
\Omega_n(t) \propto \left(-\frac{i}{\hbar}\right)^n \int_0^t dt_1 \int_0^{t_1} dt_2 \cdots \int_0^{t_{n-1}} dt_n \left[ \hat{H}(t_1), \left[ \hat{H}(t_2), \cdots, \hat{H}(t_n) \right] \cdots \right] + \text{permutations}
\end{equation}
In many cases of application we will limit to:

\begin{equation}
\Omega(t) \approx \Omega_1(t) + \Omega_2(t)
\end{equation}
The exponential will therefore be:

\begin{equation}
\exp\left(\Omega(t)\right) \approx \exp\left( \Omega_1(t) + \Omega_2(t) \right)
\end{equation}
With $\Omega_1$ the integral of $\hat{H}(t)$ and $\Omega_2$ the first corrective in commutator.
\subsection*{Quantum Production Function}

The global state of the economy at time $t$ is a quantum vector $\ket{\psi(t)}$ belonging to the Hilbert space seen upstream.

The quantum production function is defined by the expectation of the corresponding operator in the state $\ket{\psi(t)}$:

\begin{equation}
Y(t) = \bra{\psi(t)} \hat{A}(t) \hat{K}(t)^\alpha \hat{L}(t)^\beta \ket{\psi(t)}
\end{equation}
Let’s take up the eigenbases seen earlier:

\begin{itemize}
  \item $\{\ket{K_i}\}$ of $\mathcal{H}_K$, such that $\hat{K}(t)\ket{K_i} = K_i(t)\ket{K_i}$
  \item $\{\ket{L_j}\}$ of $\mathcal{H}_L$, such that $\hat{L}(t)\ket{L_j} = L_j(t)\ket{L_j}$
  \item $\{\ket{A_k}\}$ of $\mathcal{H}_A$, such that $\hat{A}(t)\ket{A_k} = A_k(t)\ket{A_k}$
\end{itemize}
These bases are assumed to be orthonormal so that:

\[
\braket{K_i | K_{i'}} = \delta_{i,i'}, \quad \braket{L_j | L_{j'}} = \delta_{j,j'}, \quad \braket{A_k | A_{k'}} = \delta_{k,k'}
\]
The operators are then written as:

\begin{align}
\hat{K}(t) &= \sum_i K_i(t) \ket{K_i} \bra{K_i}, \quad
\hat{L}(t) = \sum_j L_j(t) \ket{L_j} \bra{L_j}, \quad
\hat{A}(t) = \sum_k A_k(t) \ket{A_k} \bra{A_k}
\end{align}
The model assumes a Cobb-Douglas type production function with $\hat{K}(t)^\alpha$ and $\hat{L}(t)^\beta$. We therefore have:

\begin{align}
\hat{K}(t)^\alpha &= \sum_i [K_i(t)]^\alpha \ket{K_i} \bra{K_i} \\
\hat{L}(t)^\beta &= \sum_j [L_j(t)]^\beta \ket{L_j} \bra{L_j}
\end{align}
The operator $\hat{A}(t)$ is diagonal in $|A_k\rangle$:

\begin{equation}
\hat{A}(t) = \sum_k A_k(t) \ket{A_k} \bra{A_k}
\end{equation}
When forming the product $\hat{A}(t) \hat{K}(t)^\alpha \hat{L}(t)^\beta$, the general non-commutativity of the operators is not a problem insofar as each is diagonal in its own space. However, $K$, $L$, and $A$ act on different spaces. On the tensor product $\mathcal{H}_K \otimes \mathcal{H}_L \otimes \mathcal{H}_A$, the total operator is then written:

\begin{multline}
\hat{A}(t) \hat{K}(t)^\alpha \hat{L}(t)^\beta = 
\left(\sum_k A_k(t) \ket{A_k} \bra{A_k} \right) \\
\otimes \left(\sum_i [K_i(t)]^\alpha \ket{K_i} \bra{K_i} \right) 
\otimes \left(\sum_j [L_j(t)]^\beta \ket{L_j} \bra{L_j} \right)
\end{multline}
The quantum state $\ket{\psi(t)}$ meanwhile, can be written in the tensor basis $\{\ket{K_i} \otimes \ket{L_j} \otimes \ket{A_k}\}$:

\begin{equation}
\ket{\psi(t)} = \sum_{i,j,k} \psi_{i,j,k}(t) \ket{K_i} \otimes \ket{L_j} \otimes \ket{A_k}
\end{equation}
Where $\psi_{i,j,k}(t)$ are the complex amplitudes satisfying:

\begin{equation}
\sum_{i,j,k} |\psi_{i,j,k}(t)|^2 = 1
\end{equation}
Now let's evaluate the mean value:

\begin{equation}
Y(t) = \bra{\psi(t)} \hat{A}(t) \hat{K}(t)^\alpha \hat{L}(t)^\beta \ket{\psi(t)}
\end{equation}
Inserting the identity resolutions in the eigenbases, we obtain:

\begin{equation}
Y(t) = \sum_{i,j,k} \sum_{i',j',k'} \psi_{i',j',k'}^*(t) \bra{K_{i'} L_{j'} A_{k'}} 
\hat{A}(t) \hat{K}(t)^\alpha \hat{L}(t)^\beta 
\ket{K_i L_j A_k} \psi_{i,j,k}(t)
\end{equation}
Thanks to orthogonality, and knowing that each operator acts on its own space, we have:

\[
\hat{A}(t) \ket{A_k} = A_k(t) \ket{A_k}, \quad
\hat{K}(t)^\alpha \ket{K_i} = [K_i(t)]^\alpha \ket{K_i}, \quad
\hat{L}(t)^\beta \ket{L_j} = [L_j(t)]^\beta \ket{L_j}
\]
The action of $\hat{A}(t)\hat{K}(t)^\alpha\hat{L}(t)^\beta$ on $\ket{K_i L_j A_k}$ is therefore:

\begin{equation}
\hat{A}(t) \hat{K}(t)^\alpha \hat{L}(t)^\beta \ket{K_i L_j A_k} = 
A_k(t) [K_i(t)]^\alpha [L_j(t)]^\beta \ket{K_i L_j A_k}
\end{equation}
Substituting in the expression of $Y(t)$, the transition matrix simplifies considerably:

\begin{equation}
Y(t) = \sum_{i,j,k} \sum_{i',j',k'} \psi_{i',j',k'}^*(t)
A_k(t) [K_i(t)]^\alpha [L_j(t)]^\beta 
\delta_{i,i'} \delta_{j,j'} \delta_{k,k'} \psi_{i,j,k}(t)
\end{equation}
Thus, using the Kronecker delta contractions:

\begin{equation}
Y(t) = \sum_{i,j,k} |\psi_{i,j,k}(t)|^2 A_k(t) [K_i(t)]^\alpha [L_j(t)]^\beta
\end{equation}
The quantum production function, by taking into account the operators 
$\hat{K}(t)$, $\hat{L}(t)$, and $\hat{A}(t)$, offers an overview of the dynamic interactions between capital, labor and innovation. By quantifying these interactions, it provides a better understanding of how technological change and market fluctuations influence the economy's overall output.
The production $Y(t)$ at an instant $t$, in a quantum framework, is thus given by the average of the quantum production operator applied to the state $\ket{\psi(t)}$:

\begin{equation}
Y(t) = \bra{\psi(t)} \hat{A}(t) \cdot \hat{K}(t)^\alpha \cdot \hat{L}(t)^\beta \ket{\psi(t)}
\end{equation}
\section{Uncertainty and Fluctuations – Economic}

\subsection*{Commutation and Uncertainty}

Interactions between production variables can now be modeled by commutation relationships between their operators. These relations describe how different economic variables interact with each other in a quantum framework. Thus the commutators $[\hat{K}(t), \hat{L}(t)]$, $[\hat{K}(t), \hat{A}(t)]$, $[\hat{L}(t), \hat{A}(t)]$, can be modeled as follows:

\begin{align}
[\hat{K}(t), \hat{L}(t)] &= i\hbar \lambda \hat{L}(t) \\
[\hat{K}(t), \hat{A}(t)] &= i\hbar \alpha \hat{A}(t) \\
[\hat{L}(t), \hat{A}(t)] &= i\hbar \beta \hat{A}(t)
\end{align}
$\hbar$ is here an economic quantization constant, and $\lambda$, $\alpha$, $\beta$ are real parameters.

\subsection*{Heisenberg Principle}

Quantum uncertainty is not simply a question of measurement precision; it reflects the very nature of the economy as a dynamic system where economic variables can evolve unpredictably, influenced by a multitude of internal and external factors.
The uncertainty of the variables in the model can be modeled for each variable as follows:

\begin{align}
\Delta K(t) &= \sqrt{ \langle \hat{K}^2(t) \rangle - \langle \hat{K}(t) \rangle^2 } \\
\Delta L(t) &= \sqrt{ \langle \hat{L}^2(t) \rangle - \langle \hat{L}(t) \rangle^2 } \\
\Delta A(t) &= \sqrt{ \langle \hat{A}^2(t) \rangle - \langle \hat{A}(t) \rangle^2 }
\end{align}
We now need to describe the uncertainty relationships for each pair of key economic variables:

\begin{align}
\Delta K(t) \cdot \Delta L(t) &\geq \frac{1}{2} |\langle [\hat{K}(t), \hat{L}(t)] \rangle| \\
\Delta K(t) \cdot \Delta A(t) &\geq \frac{1}{2} |\langle [\hat{K}(t), \hat{A}(t)] \rangle| \\
\Delta A(t) \cdot \Delta L(t) &\geq \frac{1}{2} |\langle [\hat{L}(t), \hat{A}(t)] \rangle|
\end{align}
\subsection*{Complex Commutations}

In a more complex framework, commutation relations can also be extended to model tripartite interactions between capital, labor and technological innovation. This is modeled as follows:

\begin{equation}
[\hat{K}(t), [\hat{L}(t), \hat{A}(t)]] = i\hbar (\gamma \hat{L}(t) \hat{A}(t) + \delta \hat{K}(t) \hat{A}(t))
\end{equation}
This relationship describes a process in which the interactions between capital, labor and innovation are not simply independent, but are affected by feedback loops that modify the impact of each variable on the others.

Let's now insert a non-linear feedback term:

\begin{equation}
[\hat{K}(t), [\hat{L}(t), [\hat{A}(t), \hat{K}(t)]]] = i\hbar \epsilon \hat{L}(t) \hat{A}(t) \hat{K}(t)
\end{equation}
As for the final switch, it inserts the temporal dimension; let's start again from the initial switches:

\begin{align}
[\hat{K}(t), \hat{L}(t)] &= i\hbar \lambda \hat{L}(t) e^{-\gamma t} \\
[\hat{K}(t), \hat{A}(t)] &= i\hbar \alpha \hat{A}(t) e^{-\delta t} \\
[\hat{L}(t), \hat{A}(t)] &= i\hbar \beta \hat{A}(t) e^{-\eta t}
\end{align}
The factors $e^{-\gamma t}$, $e^{-\delta t}$, $e^{-\eta t}$ reflect a temporal attenuation of the intensity of non-commutative effects.
Integrating these factors into the uncertainty relations gives:

\begin{align}
\Delta K(t) \cdot \Delta L(t) &\geq \frac{1}{2} \hbar |\lambda \langle L(t) \rangle| e^{-\gamma t} \\\\
\Delta K(t) \cdot \Delta A(t) &\geq \frac{1}{2} \hbar |\alpha \langle A(t) \rangle| e^{-\delta t} \\\\
\Delta A(t) \cdot \Delta L(t) &\geq \frac{1}{2} \hbar |\beta \langle A(t) \rangle| e^{-\eta t}
\end{align}
With $\gamma > 0$, $\delta > 0$, $\eta > 0$ representing a decrease in the intensity of the effect with time.
By adding a non-linear feedback effect between the three variables, we can now model interactions that are not only affected by time but also by instantaneous feedback loops between economic variables:

\begin{equation}
[\hat{K}(t), [\hat{L}(t), [\hat{A}(t), \hat{K}(t)]]] = i\hbar \epsilon^{-\lambda t} (\gamma \hat{L}(t) \hat{A}(t) + \delta \hat{K}(t) \hat{A}(t)) e^{-\alpha t}
\end{equation}
We now need to model economic fluctuations in this switch:

\begin{equation}
[\hat{K}(t), [\hat{L}(t), [\hat{A}(t), \hat{K}(t)]]] = i\hbar \epsilon^{-\lambda t} (\gamma \hat{L}(t) \hat{A}(t) + \delta \hat{K}(t) \hat{A}(t)) e^{-\alpha t} + \eta \hat{K}(t) \hat{L}(t)
\end{equation}
At the heart of the proposed approach is the idea that the non-commutativity of economic variables reflects asymmetries and structural frictions at the microeconomic level.

Instead of arbitrarily postulating commutators, we can seek to derive them from a detailed model in which heterogeneous agents interact sequentially and non-linearly. For example, consider a set of producers who successively make capital investment ($K$) and labor recruitment ($L$) decisions.

In a conventional context, the order in which these decisions are taken may seem trivial. However, if one introduces adjustment frictions—such as non-linear hiring costs, information asymmetries, or institutional constraints preventing instantaneous adjustment—the order of decisions becomes crucial.

By aggregating the behavior of these multiple agents, a non-commutative structure emerges at the macroeconomic level, in which operators no longer commute, reflecting the irreversibility and asymmetry of decision-making processes.

From then on, the switching relations

\[
[\hat{K}(t), \hat{L}(t)], \quad [\hat{K}(t), \hat{A}(t)], \quad [\hat{L}(t), \hat{A}(t)] \neq 0
\]

cease to be arbitrary and become the direct consequence of a specific microfounded model.
\subsection*{Langevin Equation}

The Langevin equation is used to describe the evolution of an economic variable under the influence of deterministic forces, as well as random disturbances, often represented by white noise $\xi(t)$.

Consider a generalized quantum economic variable $\hat{X}(t)$—which could be $\hat{K}(t)$, $\hat{L}(t)$ or $\hat{A}(t)$—whose dynamics is affected by both a deterministic component and stochastic fluctuations.

The Langevin equation, inspired by stochastic process models in physics, is written here as:

\begin{equation}
\frac{d\hat{X}(t)}{dt} = -\gamma \hat{X}(t) + \eta \cdot \xi(t)
\end{equation}
In this equation, $\gamma > 0$ is a dissipation coefficient that controls the decay of the variable $\hat{X}(t)$ over time, while $\eta$ is the fluctuation intensity factor. $\xi(t)$ is therefore Gaussian white noise with:

\[
\langle \xi(t) \rangle = 0, \quad \langle \xi(t) \xi(t') \rangle = \delta(t - t')
\]
where $\delta(t - t')$ is Dirac's delta function.
We must first rewrite the differential equation in a standard linear first-order form:

\begin{equation}
\frac{d\hat{X}(t)}{dt} + \gamma \hat{X}(t) = \eta \xi(t)
\end{equation}
With $\gamma \hat{X}(t)$ acting as a recall term towards zero and $\eta \xi(t)$ representing a random forcing.

We now need to use the factoring integration method. The integrating factor is given by:

\[
\mu(t) = e^{\int \gamma dt} = e^{\gamma t}
\]
Multiplying the equation by $e^{\gamma t}$ gives:

\begin{equation}
e^{\gamma t} \frac{d\hat{X}(t)}{dt} + \gamma e^{\gamma t} \hat{X}(t) = \eta e^{\gamma t} \xi(t)
\end{equation}
But $e^{\gamma t} \frac{d\hat{X}(t)}{dt} + \gamma e^{\gamma t} \hat{X}(t)$ is simply the derivative of the product $e^{\gamma t} \hat{X}(t)$:

\begin{equation}
\frac{d}{dt} \left( e^{\gamma t} \hat{X}(t) \right) = \eta e^{\gamma t} \xi(t)
\end{equation}
Now let's integrate the two members of this equation between $0$ and $t$:

\begin{equation}
\int_0^t \frac{d}{dt'} \left( e^{\gamma t'} \hat{X}(t') \right) dt' = \int_0^t \eta e^{\gamma t'} \xi(t') dt'
\end{equation}
The left-hand side integral evaluates to:

\[
e^{\gamma t} \hat{X}(t) - e^{\gamma \cdot 0} \hat{X}(0) = e^{\gamma t} \hat{X}(t) - \hat{X}(0)
\]
So we get:

\begin{equation}
e^{\gamma t} \hat{X}(t) - \hat{X}(0) = \eta \int_0^t e^{\gamma t'} \xi(t') dt'
\end{equation}
Isolating $\hat{X}(t)$ we obtain:

\begin{equation}
\hat{X}(t) = \hat{X}(0) e^{-\gamma t} + \eta e^{-\gamma t} \int_0^t e^{\gamma t'} \xi(t') dt'
\end{equation}
Let’s reformulate the integral to now highlight the factor $e^{-\gamma(t - t')}$:

\begin{equation}
\hat{X}(t) = \hat{X}(0) e^{-\gamma t} + \eta \int_0^t e^{-\gamma(t - t')} \xi(t') dt'
\end{equation}
\section{Stability and Fixed Points}

In this section, we focus on analyzing the stability of the quantum economic system we have modeled so far. In particular, we study the system's fixed points, i.e., the configurations of economic variables that no longer vary over time.

Analyzing the stability of these fixed points enables us to understand how the system reacts to small perturbations, and to determine whether the economic system can converge to a stable equilibrium or is likely to diverge under specific conditions.

The economic variables $\hat{K}(t)$, $\hat{L}(t)$ and $\hat{A}(t)$ evolve over time according to differential equations that depend on their mutual interactions. We model the dynamic evolution of these variables by a system of differential equations:

\begin{align}
\frac{d\hat{K}(t)}{dt} &= f(\hat{K}(t), \hat{L}(t), \hat{A}(t)) \\
\frac{d\hat{L}(t)}{dt} &= g(\hat{K}(t), \hat{L}(t), \hat{A}(t)) \\
\frac{d\hat{A}(t)}{dt} &= h(\hat{K}(t), \hat{L}(t), \hat{A}(t))
\end{align}
The functions $f$, $g$ and $h$ capture the interactions between these variables, including the non-linear effects resulting from their mutual feedback.
The aim is to study the existence and stability of fixed points, i.e., the values of $\hat{K}(t)$, $\hat{L}(t)$ and $\hat{A}(t)$ for which the derivatives with respect to time are zero:

\begin{equation}
\frac{d\hat{K}(t)}{dt} = 0, \quad \frac{d\hat{L}(t)}{dt} = 0, \quad \frac{d\hat{A}(t)}{dt} = 0
\end{equation}
This amounts to solving the following system of equations:

\begin{align}
f(\hat{K}^*, \hat{L}^*, \hat{A}^*) &= 0 \\
g(\hat{K}^*, \hat{L}^*, \hat{A}^*) &= 0 \\
h(\hat{K}^*, \hat{L}^*, \hat{A}^*) &= 0
\end{align}
where $\hat{K}^*$, $\hat{L}^*$ and $\hat{A}^*$ are the fixed points of the system, representing the equilibrium values of the economic variables.
To analyze the stability of the fixed points, we linearize the system around $(\hat{K}^*, \hat{L}^*, \hat{A}^*)$ by calculating the Jacobian of the system of equations.
The Jacobian is the matrix of partial derivatives of the functions $f$, $g$ and $h$ with respect to the variables $\hat{K}$, $\hat{L}$ and $\hat{A}$, evaluated at the fixed point.
The Jacobian in this model is then given by:

\begin{equation}
J = 
\begin{bmatrix}
\frac{\partial f}{\partial \hat{K}} & \frac{\partial f}{\partial \hat{L}} & \frac{\partial f}{\partial \hat{A}} \\
\frac{\partial g}{\partial \hat{K}} & \frac{\partial g}{\partial \hat{L}} & \frac{\partial g}{\partial \hat{A}} \\
\frac{\partial h}{\partial \hat{K}} & \frac{\partial h}{\partial \hat{L}} & \frac{\partial h}{\partial \hat{A}}
\end{bmatrix}_{(\hat{K}^*, \hat{L}^*, \hat{A}^*)}
\end{equation}

The stability of the fixed point depends on the eigenvalues of the Jacobian.

- If all eigenvalues have negative real parts, the fixed point is \textbf{stable} and the system returns to this equilibrium after a perturbation.
- If one or more eigenvalues have positive real parts, the fixed point is \textbf{unstable} and the system diverges.

So let's take a simple example where we model the interactions between $\hat{K}(t)$, $\hat{L}(t)$ and $\hat{A}(t)$ using specific functions $f$, $g$ and $h$. Assume the following relationships:

\begin{align}
f(\hat{K}, \hat{L}, \hat{A}) &= -\alpha \hat{K} + \beta \hat{L} \hat{A} \\
g(\hat{K}, \hat{L}, \hat{A}) &= \gamma \hat{K} - \delta \hat{L} \hat{A} \\
h(\hat{K}, \hat{L}, \hat{A}) &= \epsilon \hat{L} - \zeta \hat{K} \hat{A}
\end{align}

To find the fixed points, we solve the following system:

\begin{align}
-\alpha \hat{K} + \beta \hat{L} \hat{A} &= 0 \\
\gamma \hat{K} - \delta \hat{L} \hat{A} &= 0 \\
\epsilon \hat{L} - \zeta \hat{K} \hat{A} &= 0
\end{align}
\section*{Feynman Integral Approach}

The key idea here will be to discretize the time interval $[0,t]$ into $N$ small sub-intervals, then take the limit $N \to \infty$. At each sub-interval, we insert identity resolutions that transform the exponential evolution operator into a product of simpler exponential operators.
To illustrate our approach, we consider a standard one-variable Hamiltonian:
\begin{equation}
\hat{H}(\hat{x}, \hat{p}) = \frac{\hat{p}^2}{2m} + V(\hat{x})
\end{equation}
The transition amplitude is then defined as follows:
\begin{equation}
\langle x_f | e^{-i \hat{H} t / \hbar} | x_i \rangle
\end{equation}
To obtain the exponential operator as a product we must:  
\begin{itemize}
  \item Divide the interval $[0,t]$ into $N$ segments of size $\Delta t = t/N$
  \item Define the instants $t_n = n \Delta t$ for $n = 0,1,\dots,N$. Thus, $t_0 = 0$ and $t_N = t$
  \item Set $x_0 = x_i$ and $x_N = x_f$. We then need to introduce intermediate variables $x_1, x_2, \dots, x_{N-1}$ at times $t_1, t_2, \dots, t_{N-1}$
\end{itemize}
The operator is then written:
\begin{equation}
e^{-i \hat{H} t / \hbar} \approx \underbrace{e^{-i \hat{H} \Delta t / \hbar} \cdot e^{-i \hat{H} \Delta t / \hbar} \cdots e^{-i \hat{H} \Delta t / \hbar}}_{N\text{ terms}}
\end{equation}
in the limit $\Delta t \to 0$ (thus $N \to \infty$). Between each factor $e^{-i \hat{H} \Delta t / \hbar}$ we insert the identity $I = \int |x\rangle \langle x| \, dx$. This gives:
\begin{equation}
\langle x_f | e^{-i \hat{H} t / \hbar} | x_i \rangle = \int dx_1 \int dx_2 \cdots \int dx_{N-1} \langle x_f | e^{-i \hat{H} \Delta t / \hbar} | x_{N-1} \rangle \cdots \langle x_1 | e^{-i \hat{H} \Delta t / \hbar} | x_i \rangle
\end{equation}
We thus obtain a product of $N$ matrices of the type
\begin{equation}
\langle x_n | e^{-i \hat{H} \Delta t / \hbar} | x_{n-1} \rangle
\end{equation}
with $x_0 = x_i$ and $x_N = x_f$.
Now we need to factor $\hat{p}^2 / 2m$ as well as $V(\hat{x})$. So, when $\Delta t$ is very small, we must use Trotter's formula:
\begin{equation}
e^{-i (\hat{p}^2 / 2m + V(\hat{x})) \Delta t / \hbar} \approx e^{-i \hat{p}^2 \Delta t / 2m \hbar} e^{-i V(\hat{x}) \Delta t / \hbar} + \mathcal{O}(\Delta t^2)
\end{equation}
This means factoring the exponential of a sum of non-commutative operators. The error is of the order $\Delta t^2$. For $\Delta t \to 0$, this becomes exact in the limit.
To evaluate each amplitude
\begin{equation}
\langle x_n | e^{-i \hat{H} \Delta t / \hbar} | x_{n-1} \rangle,
\end{equation}
we need to insert the momentum basis identity \( I = \int |p\rangle \langle p| \, dp \). We then obtain:
\begin{equation}
\langle x_n | e^{-i \hat{H} \Delta t / \hbar} | x_{n-1} \rangle \approx \int dp_n \, \langle x_n | e^{-i \hat{p}^2 \Delta t / 2m\hbar} | p_n \rangle \langle p_n | e^{-i V(\hat{x}) \Delta t / \hbar} | x_{n-1} \rangle.
\end{equation}
\noindent
\(\hat{p}^2/2m\) acts in the \(|p_n\rangle\) basis as:
\begin{equation}
e^{-i \hat{p}^2 \Delta t / 2m\hbar} |p_n\rangle = e^{-i p_n^2 \Delta t / 2m\hbar} |p_n\rangle.
\end{equation}
Furthermore, we have:
\begin{equation}
\langle x_n | p_n \rangle = \frac{1}{\sqrt{2\pi\hbar}} e^{i p_n x_n / \hbar}.
\end{equation}
\( \hat{V}(\hat{x}) \), on the other hand, acts diagonally in \(|x\rangle\), giving:
\begin{equation}
\langle p_n | e^{-i V(\hat{x}) \Delta t / \hbar} | x_{n-1} \rangle = e^{-i V(x_{n-1}) \Delta t / \hbar} \langle p_n | x_{n-1} \rangle = \frac{1}{\sqrt{2\pi\hbar}} e^{-i V(x_{n-1}) \Delta t / \hbar} e^{-i p_n x_{n-1} / \hbar}.
\end{equation}
Combining all the terms, we then obtain:
\begin{equation}
\langle x_n | e^{-i \hat{H} \Delta t / \hbar} | x_{n-1} \rangle 
\simeq \int \frac{dp_n}{2\pi\hbar} 
\exp\left[
\frac{i}{\hbar} p_n (x_n - x_{n-1}) 
- \frac{i}{\hbar} \frac{p_n^2}{2m} \Delta t 
- \frac{i}{\hbar} V(x_{n-1}) \Delta t
\right].
\end{equation}
When we multiply all these amplitudes for \( n = 1 \) to \( N \), as well as integrate over each \( x_1, \ldots, x_{N-1} \) and \( p_1, \ldots, p_N \), we then obtain:
\begin{equation}
\langle x_f | e^{-i \hat{H} t / \hbar} | x_i \rangle 
= \lim_{N \to \infty} \int \prod_{n=1}^{N-1} dx_n 
\int \prod_{n=1}^{N} \frac{dp_n}{2\pi\hbar} 
\exp\left[
\frac{i}{\hbar} \sum_{n=1}^{N} \left( 
p_n (x_n - x_{n-1}) 
- \left( \frac{p_n^2}{2m} + V(x_{n-1}) \right) \Delta t
\right)
\right].
\end{equation}
We then interpret the sum in the exponent as an approximation of the integral of the classical action:
\begin{equation}
\sum_{n=1}^{N} \left( 
p_n (x_n - x_{n-1}) 
- \left( \frac{p_n^2}{2m} + V(x_{n-1}) \right) \Delta t
\right)
\longrightarrow 
\int_0^t \left( p(\tau) \dot{x}(\tau) - H(x(\tau), p(\tau)) \right) d\tau,
\end{equation}
with:
\begin{equation}
\frac{x_n - x_{n-1}}{\Delta t} \approx \dot{x}(t_n), \quad
\sum_n \frac{p_n^2}{2m} \Delta t \to \int_0^t \frac{p(\tau)^2}{2m} d\tau,
\end{equation}
and similarly for the potential term \( V(x_{n-1}) \).
The reconstruction of \( \int_0^t \left[ p(\tau) \dot{x}(\tau) - H(x(\tau), p(\tau)) \right] d\tau \) 
indicates that we are moving toward the limit of a functional integral:
\begin{equation}
\langle x_f | e^{-\frac{i}{\hbar} \hat{H} t} | x_i \rangle
= \int \mathcal{D}[x(\tau)] \, \mathcal{D}[p(\tau)] \, 
\exp\left( \frac{i}{\hbar} \int_0^t \left[ p(\tau) \dot{x}(\tau) - H(x(\tau), p(\tau)) \right] d\tau \right).
\end{equation}
In the phase-space formulation, we have constructed the path integral with two dynamic fields: 
the coordinate \( x(\tau) \) and its conjugate variable \( p(\tau) \). 
The action then takes the following form:
\begin{equation}
S[x(\tau), p(\tau)] = \int_0^t d\tau 
\left[ p(\tau) \dot{x}(\tau) - \left( \frac{p(\tau)^2}{2m} + V(x(\tau)) \right) \right].
\end{equation}
The functional measure in the transition amplitude then becomes:
\begin{equation}
\int \mathcal{D}x(\tau) \, \mathcal{D}p(\tau) \, 
\exp\left( \frac{i}{\hbar} S[x(\tau), p(\tau)] \right).
\end{equation}
However, for a Hamiltonian of the form:
\begin{equation}
\hat{H} = \frac{\hat{p}^2}{2m} + V(\hat{x}),
\end{equation}
the integration over any momentum $p(\tau)$ is---in principle---Gaussian, because the kinetic term $\frac{\hat{p}^2}{2m}$ depends quadratically on $p$. We therefore wish to formally perform the integration over the variable $p(\tau)$:

\begin{equation}
\int \mathcal{D}p(\tau)\, \exp\left( \frac{i}{\hbar} \int_0^t d\tau \left[ p(\tau) \dot{x}(\tau) - \frac{p(\tau)^2}{2m} - V(x(\tau)) \right] \right).
\end{equation}
To highlight the Gaussian structure of the integral with respect to $p(\tau)$, we rearrange the part of the action that depends on $p$:

\begin{equation}
 p(\tau) \dot{x}(\tau) - \frac{p(\tau)^2}{2m} = -\frac{1}{2m} \left[ p(\tau)^2 - 2m p(\tau) \dot{x}(\tau) \right].
\end{equation}
Now we complete the square for the expression $p(\tau)^2 - 2m p(\tau) \dot{x}(\tau)$:
\begin{equation}
 p(\tau)^2 - 2m p(\tau) \dot{x}(\tau) = \left[ p(\tau) - m \dot{x}(\tau) \right]^2 - m^2 \dot{x}(\tau)^2.
\end{equation}
Hence,
\begin{equation}
p(\tau) \dot{x}(\tau) - \frac{p(\tau)^2}{2m} = -\frac{1}{2m} \left( [p(\tau) - m \dot{x}(\tau)]^2 - m^2 \dot{x}(\tau)^2 \right).
\end{equation}
\textbf{}
We can then split the action into two parts:

\begin{equation}
S[x,p] = \int_0^t d\tau \left( -\frac{1}{2m} [p(\tau) - m \dot{x}(\tau)]^2 \right) 
+ \int_0^t d\tau \left( \frac{m}{2} \dot{x}(\tau)^2 - V(x(\tau)) \right)
\end{equation}
Thus, in this exponential:
\begin{equation}
\exp\left( \frac{i}{\hbar} S[x,p] \right) 
= \exp\left( \frac{i}{\hbar} \int_0^t d\tau \left( -\frac{1}{2m} [p(\tau) - m \dot{x}(\tau)]^2 \right) \right)
\cdot
\exp\left( \frac{i}{\hbar} \int_0^t d\tau \left( \frac{m}{2} \dot{x}(\tau)^2 - V(x(\tau)) \right) \right)
\end{equation}
The first exponential then becomes Gaussian in the variable $[p(\tau) - m \dot{x}(\tau)]$. 
The Gaussian part in $p$ is integrated exactly and leads to:

\begin{equation}
\int \mathcal{D}p(\tau) \, \exp\left( \frac{i}{\hbar} S[x,p] \right)
= N \exp\left( \frac{i}{\hbar} \int_0^t d\tau \left[ \frac{m}{2} \dot{x}(\tau)^2 - V(x(\tau)) \right] \right)
\end{equation}
Once $p(\tau)$ has been eliminated, Feynman's formula for our one-dimensional demonstration becomes:

\begin{equation}
\langle x_f | e^{- \frac{i}{\hbar} \hat{H} t} | x_i \rangle
= N \int \mathcal{D}x(\tau) \; \exp\left( \frac{i}{\hbar} \int_0^t d\tau \left[ \frac{m}{2} \dot{x}(\tau)^2 - V(x(\tau)) \right] \right)
\end{equation}
We now need to generalize the multivariate development to match our model. Let's take the \((x, p)\)-dimensional approach and generalize it to three pairs \((K, p_K), (L, p_L)\) and \((A, p_A)\). We then want to evaluate the transition amplitude:

\begin{equation}
\langle K_f, L_f, A_f | e^{-\frac{i}{\hbar} \hat{H} t} | K_i, L_i, A_i \rangle
\end{equation}
where our Hamiltonian will be:

\begin{equation}
\hat{H} = \hat{H}(\hat{K}, \hat{L}, \hat{A}; \hat{p}_K, \hat{p}_L, \hat{p}_A)
\end{equation}
Returning to the Schrödinger equation we saw earlier, we know that the time evolution of our state \( |\psi(t)\rangle \) is governed by:

\begin{equation}
\hat{U}(t) = e^{-\frac{i}{\hbar} \hat{H} t}
\end{equation}
We are therefore looking for the matrix element \( \hat{U}(t) \) between two states 
\( |K_i, L_i, A_i \rangle \) and \( |K_f, L_f, A_f \rangle \). So we need to fix an integer \( N \), define \( \Delta t = t/N \). The discrete instants are \( t_n = n \Delta t \), for \( n = 0, 1, \dots, N \), with \( t_0 = 0 \) and \( t_N = t \). The exponentiation property is then:

\begin{equation}
e^{-\frac{i}{\hbar} \hat{H} t} = \lim_{N \to \infty} \left( e^{-\frac{i}{\hbar} \hat{H} \Delta t} \right)^N
\end{equation}
However, when \( N \) is finite but large, we write:

\begin{equation}
e^{-\frac{i}{\hbar} \hat{H} t} \approx \left( e^{-\frac{i}{\hbar} \hat{H} \Delta t} \right)^N
\end{equation}
Then, in the case where \( \hat{H} = \hat{T} + \hat{V} \) with \( \hat{T} \) depending on the momenta \( (\hat{p}_K, \hat{p}_L, \hat{p}_A) \) and \( \hat{V} \) depending on \( (\hat{K}, \hat{L}, \hat{A}) \), we apply Trotter's formula:

\begin{equation}
e^{-\frac{i}{\hbar} (\hat{T} + \hat{V}) \Delta t} \approx 
e^{-\frac{i}{\hbar} \hat{T} \Delta t} \, e^{-\frac{i}{\hbar} \hat{V} \Delta t} + \mathcal{O}(\Delta t^2)
\end{equation}
For small \( \Delta t \), the error is negligible and in the limit \( N \to \infty \) we recover exact equality.
To insert the configuration identity \( (K_n, L_n, A_n) \) as well as the momenta \( (p_{K,n}, p_{L,n}, p_{A,n}) \), we then write:

\begin{equation}
\langle K_f, L_f, A_f | \left( e^{-\frac{i}{\hbar} \hat{H} \Delta t} \right)^N | K_i, L_i, A_i \rangle
\end{equation}
Between each pair of multiplications, we insert the identity operator \( \hat{I} \) defined as follows:
Identity in the \( \{K, L, A\} \) basis:
\begin{equation}
\hat{I}_{(K, L, A)} = \int dK \, dL \, dA \; |K, L, A\rangle \langle K, L, A|
\end{equation}

Identity in the \( \{p_K, p_L, p_A\} \) basis:
\begin{equation}
\hat{I}_{(p_K, p_L, p_A)} = \int \frac{dp_K \, dp_L \, dp_A}{(2\pi\hbar)^3} \; |p_K, p_L, p_A\rangle \langle p_K, p_L, p_A|
\end{equation}

The aim is to ensure that, in a Trotter-type factorization 
\( e^{-i \hat{T} \Delta t / \hbar} \cdot e^{-i \hat{V} \Delta t / \hbar} \), 
we can insert \( \hat{I}_{(p_K, p_L, p_A)} \) to handle \( \hat{T} \) and \( \hat{I}_{(K, L, A)} \) to handle \( \hat{V} \).
Thus, for a single time slice, we need to calculate:

\begin{equation}
\langle K_n, L_n, A_n | e^{-\frac{i}{\hbar} \hat{H} \Delta t} | K_{n-1}, L_{n-1}, A_{n-1} \rangle
\end{equation}
Then, if \( \hat{H} = \hat{T} + \hat{V} \), we write:

\begin{equation}
e^{-\frac{i}{\hbar} \hat{H} \Delta t} \approx e^{-\frac{i}{\hbar} \hat{T} \Delta t} \cdot e^{-\frac{i}{\hbar} \hat{V} \Delta t}
\end{equation}
Let’s now insert \( \hat{I}_{(p_K, p_L, p_A)} \), giving:

\begin{equation}
\langle K_n, L_n, A_n | e^{-\frac{i}{\hbar} \hat{T} \Delta t} | p_{K,n}, p_{L,n}, p_{A,n} \rangle 
\cdot 
\langle p_{K,n}, p_{L,n}, p_{A,n} | e^{-\frac{i}{\hbar} \hat{V} \Delta t} | K_{n-1}, L_{n-1}, A_{n-1} \rangle
\end{equation}
We now integrate over \( dp_{K,n} \), and use the action of \( \hat{T} \) in the \( |p_K, p_L, p_A\rangle \) basis, 
the action of \( \hat{V} \) in the \( |K, L, A\rangle \) basis, 
and the Fourier transform:

\begin{equation}
\langle K | p_K \rangle = \frac{1}{\sqrt{2\pi\hbar}} e^{+i p_K K / \hbar}
\end{equation}
In this way, for a step \( \Delta t \) we obtain a factor of the type:

\begin{align}
&\int \frac{dp_{K,n} \, dp_{L,n} \, dp_{A,n}}{(2\pi\hbar)^3} \; \times \nonumber\\
&\exp \left\{ \frac{i}{\hbar} \left[ 
p_{K,n} (K_n - K_{n-1}) + p_{L,n} (L_n - L_{n-1}) \right. \right. \nonumber\\
&\left. \left. + p_{A,n} (A_n - A_{n-1}) 
- T(p_{K,n}, p_{L,n}, p_{A,n}) \Delta t 
- V(K_{n-1}, L_{n-1}, A_{n-1}) \Delta t 
\right] \right\}
\end{align}
For \( \left( e^{- \frac{i}{\hbar} \hat{H} \Delta t} \right)^N \), 
we multiply all these factors for \( n = 1, \dots, N \), 
introducing \( (N-1) \) integrals over configuration variables and \( N \) over momentum variables:

\begin{align}
\langle K_f, L_f, A_f | e^{- \frac{i}{\hbar} \hat{H} t} | K_i, L_i, A_i \rangle
&= \lim_{N \to \infty}
\int \prod_{n=1}^{N-1} dK_n \, dL_n \, dA_n 
\int \prod_{n=1}^{N} \frac{dp_{K,n} \, dp_{L,n} \, dp_{A,n}}{(2\pi\hbar)^3} \nonumber\\
&\quad \times \exp \left\{ \frac{i}{\hbar} \sum_{n=1}^{N} \left[
p_{K,n} (K_n - K_{n-1}) 
+ p_{L,n} (L_n - L_{n-1}) 
+ p_{A,n} (A_n - A_{n-1}) \right. \right. \nonumber\\
&\quad \left. \left. 
- H(p_{K,n}, p_{L,n}, p_{A,n}; K_{n-1}, L_{n-1}, A_{n-1}) \, \Delta t
\right] \right\}
\end{align}
We then identify:
\begin{equation}
\frac{\Delta K_n}{\Delta t} \approx \dot{K}(t_n), \quad 
\frac{\Delta L_n}{\Delta t} \approx \dot{L}(t_n), \quad 
\frac{\Delta A_n}{\Delta t} \approx \dot{A}(t_n)
\end{equation}
Henceforth, when we take the limit \( \Delta t \to 0 \), 
the discrete sums over \( K, L, A \) become, for each variable:
\begin{align}
\sum_{n=1}^{N} p_{K,n} \Delta K_n &\longrightarrow \int_0^t p_K(\tau) \dot{K}(\tau) \, d\tau \\
\sum_{n=1}^{N} p_{L,n} \Delta L_n &\longrightarrow \int_0^t p_L(\tau) \dot{L}(\tau) \, d\tau \\
\sum_{n=1}^{N} p_{A,n} \Delta A_n &\longrightarrow \int_0^t p_A(\tau) \dot{A}(\tau) \, d\tau
\end{align}
Moreover, the sum over the Hamiltonian becomes:
\begin{equation}
\sum_{n=1}^{N} H(\dots) \, \Delta t 
\longrightarrow 
\int_0^t H(K(\tau), L(\tau), A(\tau); p_K(\tau), p_L(\tau), p_A(\tau)) \, d\tau
\end{equation}
We then define the action in phase space:
\begin{equation}
S = \int_0^t d\tau \left[
p_K(\tau) \dot{K}(\tau) + 
p_L(\tau) \dot{L}(\tau) + 
p_A(\tau) \dot{A}(\tau) - 
H(K, L, A; p_K, p_L, p_A)
\right]
\end{equation}
Our discrete measure
\[
\prod_{n=1}^{N} \frac{dp_{K,n} \, dp_{L,n} \, dp_{A,n}}{(2\pi\hbar)^3} 
\prod_{n=1}^{N-1} dK_n \, dL_n \, dA_n
\]
becomes, in the continuous limit \( \Delta t \to 0 \), a functional measure:
\begin{equation}
\mathcal{D}p_K(\tau) \, \mathcal{D}p_L(\tau) \, \mathcal{D}p_A(\tau) \,
\mathcal{D}K(\tau) \, \mathcal{D}L(\tau) \, \mathcal{D}A(\tau)
\end{equation}
Putting all the factors together and taking the limit \( N \to \infty \), we end up with:
\begin{equation}
\langle K_f, L_f, A_f | e^{-i \hat{H} t / \hbar} | K_i, L_i, A_i \rangle
= \int \mathcal{D}[K, L, A, p_K, p_L, p_A] \;
\exp\left( \frac{i}{\hbar} S(K, L, A; p_K, p_L, p_A) \right)
\end{equation}
If:
\begin{equation}
\hat{H} = \frac{\hat{p}_K^2}{2m_K} + \frac{\hat{p}_L^2}{2m_L} + \frac{\hat{p}_A^2}{2m_A} + V(\hat{K}, \hat{L}, \hat{A}),
\end{equation}
then, in phase space, we have:
\begin{equation}
H(K, L, A; p_K, p_L, p_A) = \frac{p_K^2}{2m_K} + \frac{p_L^2}{2m_L} + \frac{p_A^2}{2m_A} + V(K, L, A).
\end{equation}
Let’s take the specific term in \( K \):
\begin{align}
p_K(\tau) \dot{K}(\tau) - \frac{p_K(\tau)^2}{2m_K} 
&= -\frac{1}{2m_K} \left[ p_K(\tau)^2 - 2m_K p_K(\tau) \dot{K}(\tau) \right] \\
&= -\frac{1}{2m_K} \left( [p_K(\tau) - m_K \dot{K}(\tau)]^2 - m_K^2 \dot{K}(\tau)^2 \right) \\
&= -\frac{1}{2m_K} [p_K(\tau) - m_K \dot{K}(\tau)]^2 + \frac{m_K}{2} \dot{K}(\tau)^2.
\end{align}
We do the same for \( L \) and \( A \):
\begin{align}
p_L(\tau) \dot{L}(\tau) - \frac{p_L(\tau)^2}{2m_L} 
&= -\frac{1}{2m_L} [p_L(\tau) - m_L \dot{L}(\tau)]^2 + \frac{m_L}{2} \dot{L}(\tau)^2, \\
p_A(\tau) \dot{A}(\tau) - \frac{p_A(\tau)^2}{2m_A} 
&= -\frac{1}{2m_A} [p_A(\tau) - m_A \dot{A}(\tau)]^2 + \frac{m_A}{2} \dot{A}(\tau)^2.
\end{align}
The Gaussian parts 
\[
\exp\left[-\frac{i}{\hbar} \int_0^t \frac{1}{2m_K} [p_K - m_K \dot{K}]^2 d\tau \right] \]
and similarly for \( p_L \) and \( p_A \), integrate to give a functional determinant essentially independent of the trajectory \(\{K, L, A\}\). The terms \( \frac{m_K}{2} \dot{K}^2 \), \( \frac{m_L}{2} \dot{L}^2 \), and \( \frac{m_A}{2} \dot{A}^2 \) then remain in the action, while the potential \( V(K, L, A) \) appears with a negative sign.
The result is then:
\begin{align}
\langle K_f, L_f, A_f | e^{-i \hat{H} t / \hbar} | K_i, L_i, A_i \rangle
&= \int \mathcal{D}K(\tau) \, \mathcal{D}L(\tau) \, \mathcal{D}A(\tau) \; \times \nonumber \\
&\quad \exp\left( \frac{i}{\hbar} \int_0^t d\tau \left[
\frac{m_K}{2} \dot{K}^2 + \frac{m_L}{2} \dot{L}^2 + \frac{m_A}{2} \dot{A}^2 - V(K, L, A)
\right] \right)
\end{align}
In Lagrangian version: 
\begin{equation}
\int \mathcal{D}K(\tau) \, \mathcal{D}L(\tau) \, \mathcal{D}A(\tau) \;
\exp\left[ \frac{i}{\hbar} \int_0^t d\tau 
\left( 
\frac{m_K}{2} \dot{K}^2 + \frac{m_L}{2} \dot{L}^2 + \frac{m_A}{2} \dot{A}^2 
- V(K, L, A)
\right) 
\right] \cdot (\text{normalization factor})
\end{equation}

\section{Conclusion and perspectives}

\begin{figure}[H]
    \centering
    \includegraphics[width=0.8\linewidth]{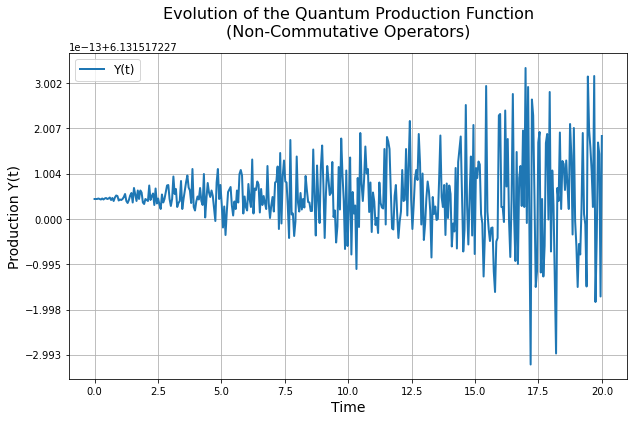}
    \caption{Evolution of the quantum production function}
    \label{fig:enter-label}
\end{figure}
The first graph in this analysis illustrates the time evolution of the quantum production function \( Y(t) \). The influence of non-commutative operators generates marked fluctuations, resulting in highly unstable and erratic dynamics. This instability is characteristic of quantum systems, where interference and superposition effects introduce non-deterministic variability into the evolution of production. \( Y(t) \) shows an amplification of oscillations as time progresses, suggesting an emerging chaotic regime.
\begin{figure}[H]
    \centering
    \begin{subfigure}[t]{0.7\linewidth}
    \includegraphics[width=0.8\linewidth]{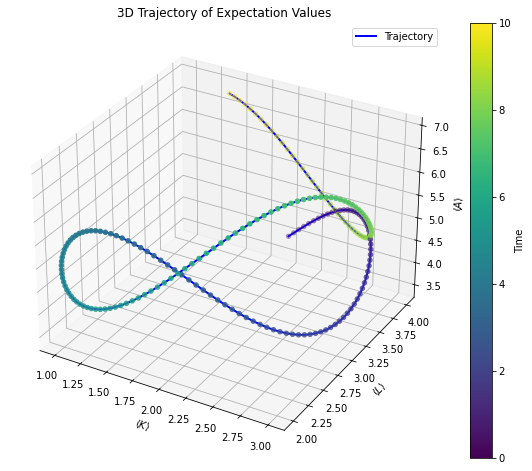}
    \caption{Trajectory of expectation values}
    \label{fig:enter-label}
    \end{subfigure}
\vspace{-0.2em}
\begin{subfigure}[t]{0.7\linewidth}
    \centering
    \includegraphics[width=0.8\linewidth]{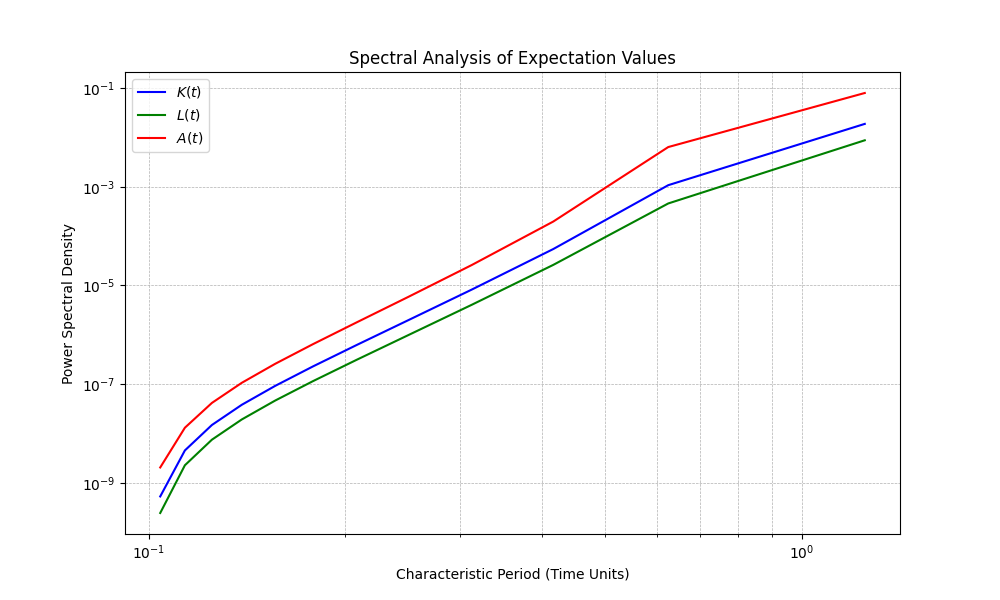}
    \caption{Spectral analysis of expectation values}
    \label{fig:enter-label}
    \end{subfigure}
    \caption{Quantum dynamics: trajectory and spectral analysis}
\end{figure}

A study of the temporal evolution of the expectation values reveals that the system initially tends towards a stabilization regime before exhibiting fluctuations of greater amplitude. This phenomenon can be interpreted as a transition to chaotic dynamics under certain parametric conditions. The three-dimensional visualization of trajectories in the space of expectation variables highlights the presence of complex attractors, suggesting that the system evolves according to a non-trivial dynamic topology, with potentially multistable behaviors.

Spectral analysis is used here to shed further light on the structure of the fluctuations observed. The accumulation of spectral density at low frequencies indicates the existence of persistent cycles, while energy dispersion at higher frequencies reflects instability and rapid transitions between quantum states. This observation is reinforced by the study of uncertainties, which gradually diminish for some variables while remaining significant for others, revealing a differentiation in the stabilization dynamics of the underlying processes.

\begin{figure}[H]
    \centering
    \includegraphics[width=0.8\linewidth]{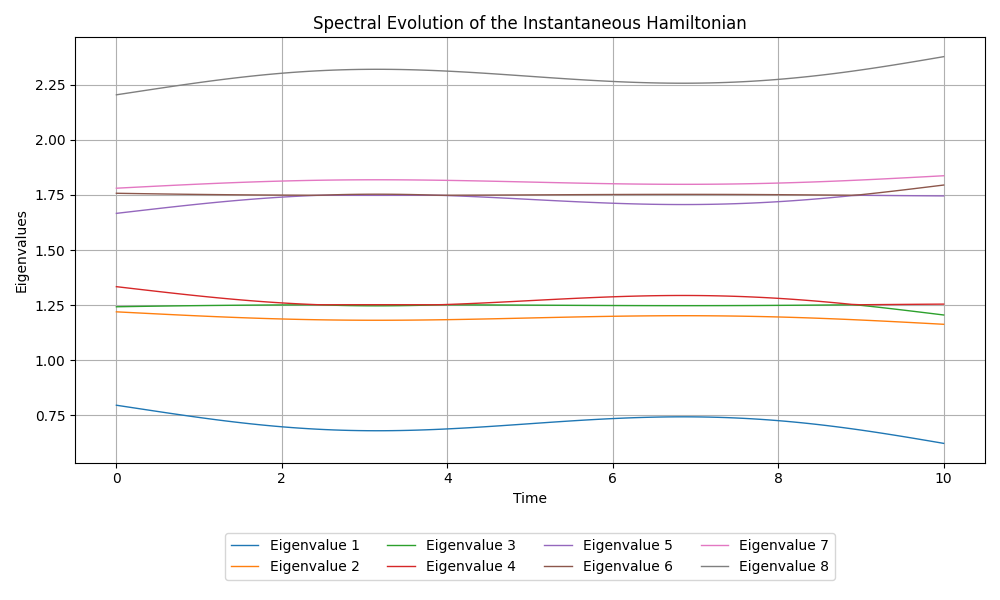}
    \caption{Spectral evolution the instantaneous Hamiltonian}
    \label{fig:enter-label}
\end{figure}
The evolution of the eigenvalues of this Hamiltonian makes it possible to analyze the underlying energy structure of the system. The existence of several clearly differentiated levels indicates a progressive separation of dynamic modes, corroborating the hypothesis of a quantum regime structured into different unstable equilibrium states. This spectral analysis highlights the presence of non-trivial couplings between system variables, directly affecting the dispersion of energy levels and influencing the trajectory of economic production in phase space.

\begin{figure}[H]
    \centering
    \includegraphics[width=0.8\linewidth]{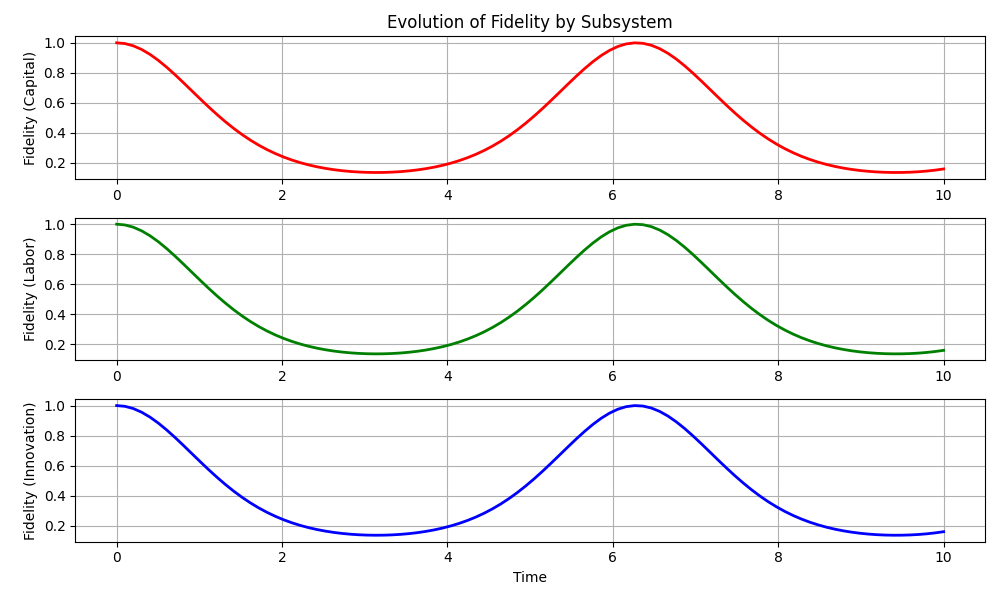}
    \caption{Evolution of fielity by subsytsem}
    \label{fig:enter-label}
\end{figure}
Quantum fidelity measures the stability of states over time, by assessing their proximity to an initial reference. The observation of marked oscillations in each subsystem reveals a fluctuating coherence of these variables, indicating a significant non-linear interaction between capital, labor and innovation. A progressive decrease in fidelity in certain phases of the cycle may suggest a phenomenon of decoherence, where the influence of non-commutative operators accentuates the divergence of individual trajectories.

\begin{figure}[H]
    \centering
    \includegraphics[width=1\linewidth]{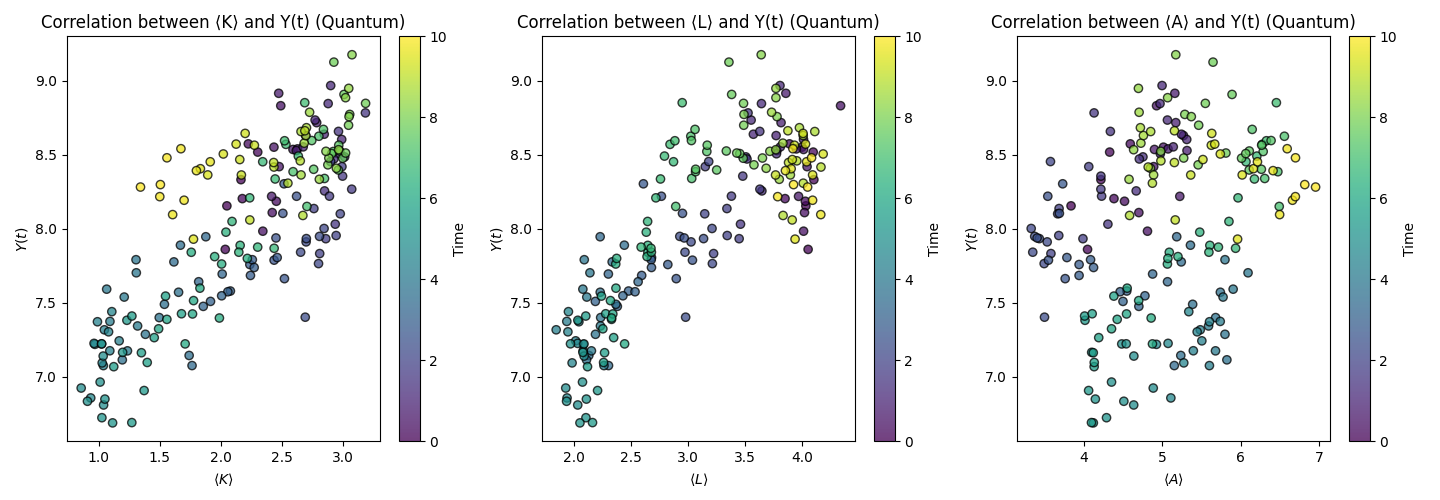}
    \caption{Correlation between variables}
    \label{fig:enter-label}
\end{figure}
Here the results are presented in sub-graphs for each expectation dimension $\langle K \rangle$, $\langle L \rangle$ and $\langle A \rangle$. In the first subplot, the correlation between $Y(t)$ and $\langle K \rangle$ highlights an increasing relationship, indicating that the evolution of quantum production is positively influenced by capital in this theoretical framework. However, the dispersion of the points suggests that other factors contribute to fluctuations in $Y(t)$. 

The second part, which links $Y(t)$ and $\langle L \rangle$, also reveals a positive trend, albeit a more heterogeneous one. The existence of different regions of concentration in state space suggests the presence of distinct regimes in the evolution of labor dynamics, which could be interpreted as differential coupling effects between subsystems. 

Finally, the third sub-graph analyzes the relationship between $Y(t)$ and $\langle A \rangle$ where we observe a correlation that strengthens as time evolves. Innovation dynamics thus seem to play a structuring role in quantum production, although non-linearities appear in certain regions of the diagram, suggesting more complex effects linked to the interaction between variables.

\begin{figure}[H]
    \centering
    \includegraphics[width=1\linewidth]{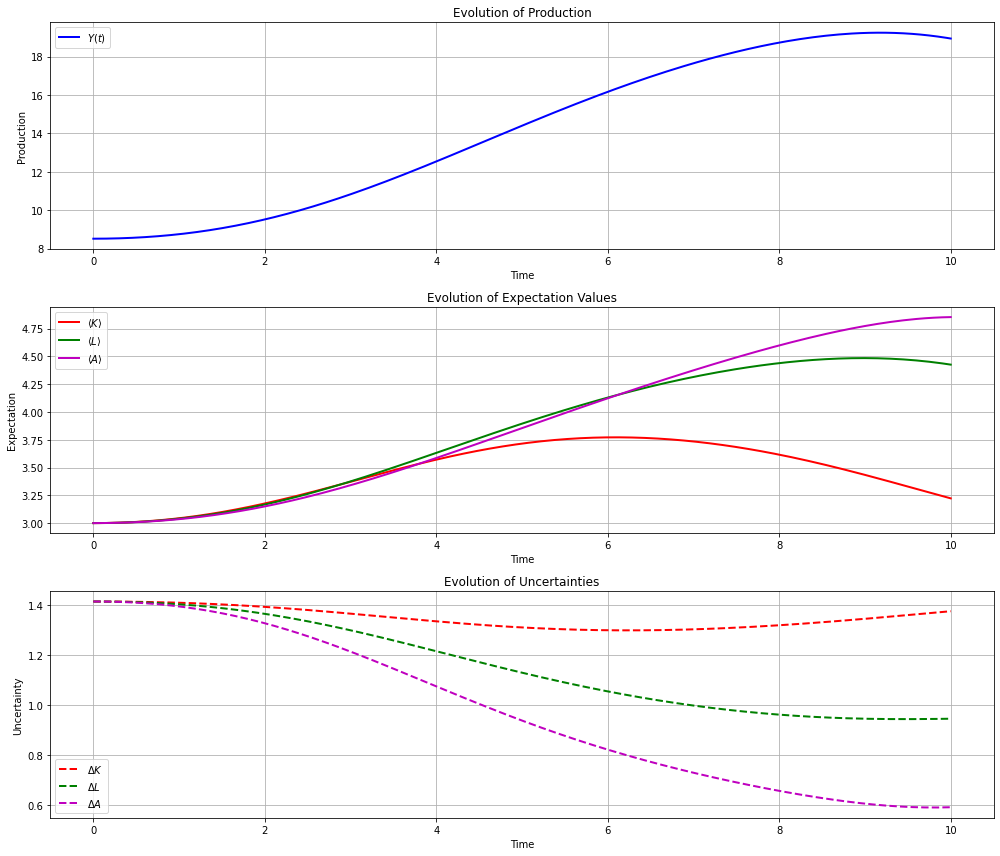}
    \caption{Evolution of the quantum system}
    \label{fig:enter-label}
\end{figure}
These observations have several implications. Firstly, they suggest that the introduction of quantum formalisms into the modelling of complex dynamics enables us to obtain a more faithful representation of phenomena where interactions are not simply additive, but incorporate interference and superposition effects. Secondly, they highlight the need for further study of dynamical transitions and bifurcations within this framework, in particular by precisely characterizing the conditions for the emergence of chaotic behavior and quantum attractors.

\begin{figure}[H]
    \centering
    \includegraphics[width=0.8\linewidth]{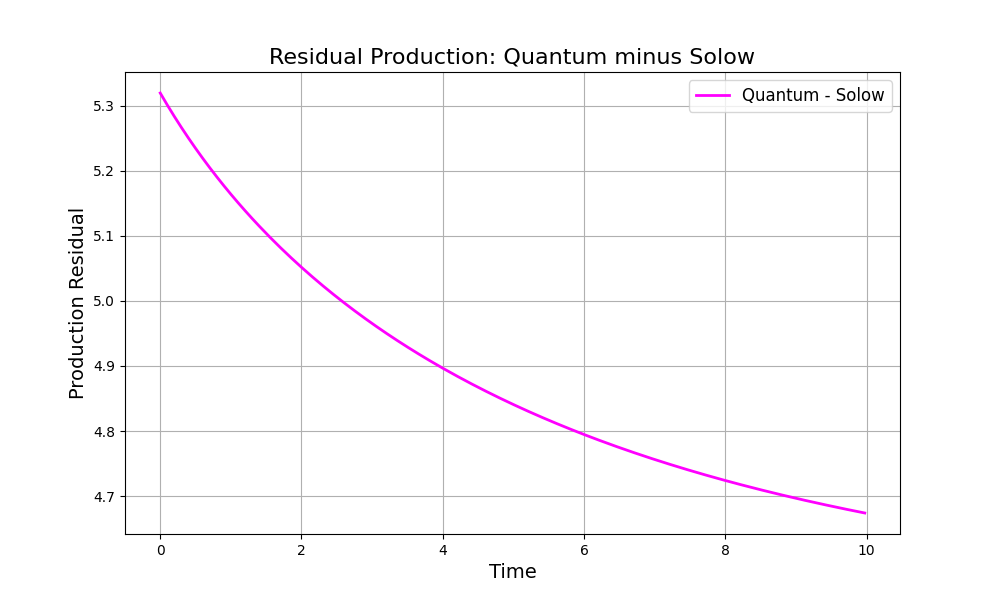}
    \caption{Residual production between quantum and Solow}
    \label{fig:enter-label}
\end{figure}
Finally, the last graph shows the production residual between the quantum model and the classical Solow model. The difference between the two dynamics shows a progressive decrease in the residual, suggesting that the quantum model provides a significant correction to classical predictions of economic growth. 
This divergence can be interpreted as the manifestation of quantum interference and correction effects introduced by the non-commutative approach. The slow decay of the residual suggests that, in the long term, classical and quantum predictions partially converge, although discrepancies persist, notably due to the non-linear dynamic effects specific to the quantum framework.

\end{document}